\documentclass[%
 superscriptaddress,
 reprint,
 amsmath,
 amssymb,
 prl,
 showpacs,
 twocolumn, 
 floatfix,
 nofootinbib,
]{revtex4-2}

\usepackage{graphicx}
\usepackage{float}
\usepackage{dcolumn}
\usepackage{bm}
\usepackage[colorlinks=true, allcolors=blue]{hyperref}
\usepackage[mathlines]{lineno}
\usepackage{color}
\usepackage[normalem]{ulem}

\begin{document}

\title{SPLENDAQ: A Detector-Agnostic Data Acquisition System for Small-Scale Physics Experiments}

\author{S. L.~Watkins} \email{slw@lanl.gov} \affiliation{Los Alamos National Laboratory, Los Alamos, NM 87545, USA}

\date{\today}

\begin{abstract}
Many scientific applications from rare-event searches to condensed matter system characterization to high-rate nuclear experiments require time-domain triggering on a raw stream of data, where the triggering is generally threshold-based or randomly acquired. When carrying out detector R\&D, there is a need for a general data acquisition (DAQ) system to quickly and efficiently process such data. In the SPLENDOR collaboration, we are developing the Python-based SPLENDAQ package for this exact purpose---it offers two main features for offline analysis of continuous data: a threshold-triggering algorithm based on the time-domain optimal filter formalism and an algorithm for randomly choosing nonoverlapping segments for noise measurements. Combined with the commercially available Moku platform, developed by Liquid Instruments, we have a full pipeline of event building off raw data with minimal setup. Here, we review the underlying principles of this detector-agnostic DAQ package and give concrete examples of its utility in various applications.
\end{abstract}

\maketitle

\textit{Introduction.---}Data acquisition (DAQ) systems are a necessary part of any scientific experiment. While large-scale experiments generally require custom systems optimized for the high data rate or complex nature of the setup~\cite{orca2004,CMS:2016ngn,DiDomizio:2018ldc,Wilson:2022quj}, small-scale experiments with few channels can benefit from a ready-made DAQ system that requires minimal setup. With large data-storage options available, an excellent strategy for acquiring data is to read out and save data from an experimental system continuously, and to then apply triggering algorithms after the fact on this saved data. There are two main triggering algorithms of interest to any experiment: randomly-acquired data and threshold-based triggering. In the SPLENDOR (Search for Particles of Light Dark Matter with Narrow-gap Semiconductors) collaboration, we have developed a Python-based DAQ system called SPLENDAQ~\cite{splendaq}, a general system that includes these algorithms for use in a variety of physics contexts.

With SPLENDAQ, the main requirement to use the provided triggering algorithms is that the user will supply a continuous stream of data, defined as large chunks (e.g. significantly longer than the expected signal) of time-series data. For example, if an experiment is searching signals with a characteristic time constant of $\mathcal{O}(1\,\mathrm{ms})$, then time-series data of length $\mathcal{O}(1\,\mathrm{s})$ should be sufficiently long enough to be thought of as a ``continuous" data stream. In this work, we describe the underlying algorithms for random and threshold triggering, the required data format to use SPLENDAQ, show example usage of the workflow, and then discuss the outlook of this package.

\textit{Data triggering algorithms.---}With a continuous data stream $v(t)$, threshold triggering algorithms can follow a multitude of strategies, such as simple tracking of values above threshold, trapezoidal filtering~\cite{JORDANOV1994261}, filtering using a known signal template~\cite{Wilson:2022quj}, etc. For SPLENDAQ, we operate under the assumption that the expected signal is of a known shape and the underlying noise of the system is approximately stationary. In this scenario, optimal filtering~\cite{radeka1967,Gatti:1986cw,golwala, Watkins:2022wlz} (also referred to as matched filtering) is frequently the strategy of choice for extracting signal amplitudes. Optimal filtering requires a known signal template $s(t)$ and a power spectral density (PSD) $J(f)$, which can be defined as the Fourier transform of the autocorrelation function of an infinitely long stream of data. 
We note that the two assumptions (known signal shape and stationary noise) are not a necessity for optimal filtering, as the algorithms within this package can still be used in cases of variation of signal shape or nonstationary noise---this may however result in a loss of energy resolution when reconstructing the signal. To calculate the PSD, a large sample of signal-free sections of the data should be acquired, which can be done via the random triggering algorithm provided in SPLENDAQ. This large sample can then be used in a PSD calculation algorithm, the result of which describes the underlying noise of the system.

The random triggering algorithm requires two inputs: the number of sections $n$ of the continuous data desired and the length $l$ of these traces in number of time bins (or samples). The underlying algorithm relies on the \textsc{NumPy}~\cite{harris2020array} algorithm \texttt{numpy.random.choice}. For an array of continuous data of shape $(m, n_\mathrm{chan}, p)$, where $m$ is the number of continuous data traces, $n_\mathrm{chan}$ is the number of channels being read out, and $p$ is the length of each continuous data chunk in time bins, the algorithm calculates the maximum number of non-overlapping sections of length $l$ that the data can be split into (i.e. $\mathrm{floor}(mp/l)$) and returns an error if the number of desired sections is greater than that number. If it is possible to extract the desired number of non-overlapping sections, then the algorithm will randomly choose indices from the array with a minimum spacing defined by $l$ without replacement and return a three-dimensional array of shape $(n, n_\mathrm{chan}, l)$. The resulting array of data can then be cleaned to remove any events that are contaminated with signals, transient noise sources, or any other phenomena that would prevent this sample from describing the underlying baseline noise of the system. The PSD can then be calculated and used in an optimal filter--based threshold triggering algorithm.

The optimal filter formalism can be explained as the least-squares minimization of a frequency-weighted $\chi^2$ between the expected pulse shape and some measured data. Because this minimization is done in frequency space, we define here the Fourier transform pair (for some function $g$) for the continuous case as
\begin{align}
    \tilde{g}(f) &= \int_{-\infty}^\infty \mathop{dt} g(t) \mathrm{e}^{-i \omega t}, \\
    g(t) &= \int_{-\infty}^\infty \mathop{df} \tilde{g}(f) \mathrm{e}^{i \omega t},
\end{align}
where $\omega \equiv 2 \pi f$ for convenience, and for the discrete case as
\begin{align}
    \tilde{g}_n &= \frac{1}{N} \sum_{k = -N/2}^{N/2 - 1}g_k  \mathrm{e}^{-i \omega_n t_k}, \\
    g_k  &= \sum_{n = -N/2}^{N/2 - 1} \tilde{g}_n \mathrm{e}^{i \omega_n t_k}.
\end{align}
In this work, we will derive equations in the continuous approximation for readability, knowing that all calculations are done in the discrete case in SPLENDAQ.

While we have already defined $v(t)$ and $s(t)$, we further define $A$ as the true height of the signal and $t_0$ as the start time of that signal. The minimization of
\begin{equation}
    \chi^2 = \int_{-\infty}^\infty \mathop{df} \frac{\left| \tilde{v}(f) - A \mathrm{e}^{-i \omega t_0} \tilde{s}(f) \right|^2}{J(f)}
\end{equation}
with respect to $A$ and $t_0$ provides the best estimate of these two quantities. Minimizing the $\chi^2$ with respect to $A$, we have that the best-fit amplitude at time $t_0$ is
\begin{equation}
    A(t_0) = \int_{-\infty}^\infty \mathop{df} \tilde{\phi}(f) \tilde{v}(f) \mathrm{e}^{-i \omega t_0}, \label{eq:ampl_freq}
\end{equation}
where we define
\begin{equation}
    \tilde{\phi}(f) \equiv \frac{\tilde{s}^*(f) / J(f)}{\int_{-\infty}^\infty \mathop{df'} \left| \tilde{s}(f') \right|^2 / J(f')}. \label{eq:chi2_freq}
\end{equation}
Equations~(\ref{eq:ampl_freq}) and (\ref{eq:chi2_freq}) define the optimal filter (see again Refs.~\cite{radeka1967,Gatti:1986cw,golwala, Watkins:2022wlz}), where the time $t_0$ at which $A(t_0)$ is maximized corresponds to the lowest $\chi^2$. The expected variance in $A$ in this formalism is calculated via the general result of $\sigma_A^2 = \left[\frac{1}{2} \frac{\partial^2 \chi^2}{\partial A^2} \right]^{-1}$, giving
\begin{equation}
    \sigma_A^2 = \left[ \int_{-\infty}^\infty \mathop{df} \frac{\left| \tilde{s}(f) \right|^2}{J(f)} \right]^{-1}.
\end{equation}
In SPLENDAQ, we convert Eq.~(\ref{eq:ampl_freq}) back to time domain and arrive at
\begin{equation}
    A(t_0) = \int_{-\infty}^\infty \mathop{dt} \phi(t - t_0)v(t),
    \label{eq:of_res}
\end{equation}
which we call the time-domain optimal filter. In this form, the optimal filter is a cross-correlation of our filter $\phi(t)$ with the time-stream data $v(t)$. In discrete space, this can be calculated as an FIR filter, where $\phi$ represents the filter coefficients being correlated with the time-stream data. In SPLENDAQ, this computation is done efficiently using \texttt{scipy.signal.correlate}~\cite{2020SciPy-NMeth}. At the edges of the data, there is ambiguity in how to handle the cross-correlation when the FIR filter coefficients do not fully overlap with the data. In these regions, we set the amplitudes of the filtered data to zero, such that there is a built-in dead time in each chunk of continuous data that is equal to the length of the FIR filter coefficients.

\begin{figure}
    \centering
    \includegraphics[width=1.0\linewidth]{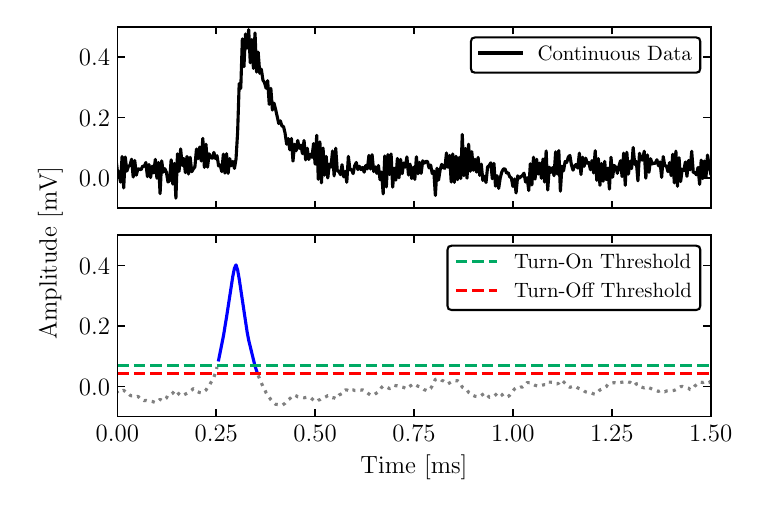}
    \caption{(Top) A section of continuous data with an event. (Bottom) A visual example of what ranges of values will be tagged as above threshold (in blue) and below threshold (in grey) in the optimal filtered data stream to be used to build events. Here, we have a $5\sigma_A$ turn-on threshold, such that the turn-off threshold is then set to be $3\sigma_A$.}
    \label{fig:turn-off}
\end{figure}

With the filtered data, we then apply a simple threshold algorithm to tag sections of data with energies reconstructed above the specified level. As we are using the optimal filter formalism, we can define this threshold in terms of number of $\sigma_A$, which provides an intuitive sense of how many noise-related triggers we can expect. When the filtered data stream goes above threshold, the algorithm tags the starting index. The end index is tagged when the filtered data stream drops below the number of $\sigma_A$ minus 2, i.e. if the turn-on threshold is $n\sigma_A$, then the turn-off threshold is $(n - 2)\sigma_A$. In Fig.~\ref{fig:turn-off}, we show an example of this algorithm on simulated data that has been optimal filtered, as compared to the continuous data stream. This hysteretic threshold method is done to reduce multiple triggers for events that are reconstructed to have energies near threshold, as these events may have fluctuations around the turn-on threshold. This case is especially important for low thresholds (e.g. $5\sigma_A$) due to the lower signal-to-noise ratio. For each range of start and end index tagged to be above threshold, the algorithm then finds the point of maximum amplitude (which corresponds one-to-one to the minimum $\chi^2$) and saves a waveform that is centered at this point (of the same length as the provided signal template). If the turn-on threshold is set to be lower than $5\sigma_A$, then the turn-off threshold is set to $3\sigma_A$ to avoid having the turn-off threshold so low that many events may be merged and treated as single events. In the case that the turn-on threshold is set to below $3\sigma_A$ (which may be unwise for both a high trigger rate and the likelihood of merging distinct events into one), then the turn-off threshold is identical to the turn-on threshold. The turn-off threshold is also specifiable by the user, if a different threshold is desired than the default behavior outlined above.

\begin{figure}
    \centering
    \includegraphics[width=1.0\linewidth]{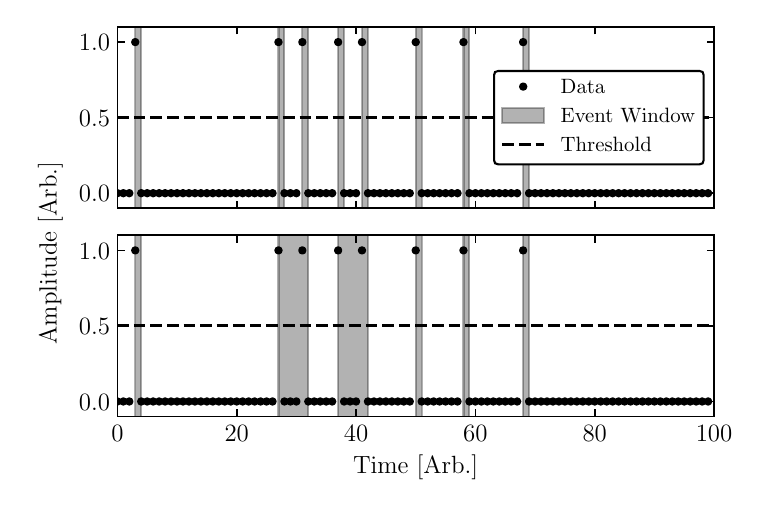}
    \caption{In this illustration of the event merging algorithm, we show a stream of data of zeroes and ones, where a threshold is set at 0.5. In the top figure, we show the event windows that have been marked as above threshold without any event merging being applied. In the bottom figure, we have set a merging window of 5 time bins, and we can see that the consecutive events between about 20 and 50\,s have been merged, as they were $\leq5$ time bins apart.}
    \label{fig:merge}
\end{figure}
There is also the possibility that there are multiple events that have been tagged above threshold that are closely spaced together (e.g. within the length of a waveform). Depending on what is desired by the user, the algorithm allows for the definition of a merging window, which by default is not used (i.e. multiple events above threshold that are arbitrarily close together remain marked as separate events). If a merging window is specified, then the algorithm checks if there are any ranges of indices that have been tagged as above threshold that are spaced by this merging window (or closer). If so, then these events are merged by redefining the indices above threshold as the start index of the first event above threshold and the end index as that of last event above threshold within this merging window. We note that if there is a long bunch of events being merged, such that the total length is longer than the requested trace length, then the algorithm centers on the section with the largest amplitude and may miss events outside the trace length. In Fig.~\ref{fig:merge}, we provide an illustration of this event merging algorithm.

\begin{table*}
    \caption{\label{tab:dataformat}Fields needed in the HDF5 file format to be used in SPLENDAQ.}
      \centering
        \begin{tabular}{lll} \hline\hline
        Field Name & Short Description & Data Type \\ \hline
          \texttt{data} & Array containing the raw data to be processed & 3D array of floats \\
          \texttt{channels}   & The names of the channels (second dimension of \texttt{data})& 1D array of strings \\
          & dimension of \texttt{data})& of strings \\
          \texttt{comment}   & Any comments on the data can be placed here & string \\
          \texttt{fs}   & Digitization rate of the data in units of Hz & float \\
          \texttt{datashape}   & Shape of \texttt{data} (number of waveforms, number of channels, length of waveforms)& 3D array of integers \\
          \texttt{eventindex}   & The index at which the event starts within the original data stream & 1D array of integers \\
          \texttt{eventnumber} & A sequential number that counts the order of events as they are tagged & 1D array of integers \\
          \texttt{eventtime}   & The time of the event in epoch time & 1D array of floats \\
          \texttt{seriesnumber}   & The date and time the data was taken (formatted as \texttt{YYYYMMDDhhmmss})& 1D array of integers\\
          \texttt{dumpnumber}  & If the data has been split into multiple files, an integer specifying the file number& 1D array of integers\\
          \texttt{triggertime}   & The time at which the trigger occurs in epoch time & 1D array of floats \\
          \texttt{triggertype}  & The type of trigger (0 = randoms, 1 = threshold) &  1D array of integers \\
          \texttt{parentseriesnumber}   & If built from a continuous data file, the  \texttt{seriesnumber} of the original file & 1D array of integers \\
          \texttt{parenteventnumber}   & If built from a continuous data file, the \texttt{eventnumber} of the original event & 1D array of integers \\ \hline\hline
        \end{tabular}
\end{table*}

\textit{Data format requirements.---}In order to use SPLENDAQ on any raw continuous stream of data, the raw data must be saved to an HDF5~\cite{hdf5} file with a specific format. In Table~\ref{tab:dataformat}, we detail the fields needed, short definitions, and the data type of each field. These metadata allow for the unique identification of events extracted from data, as well as the ability to backtrack events to the continuous data stream for further analysis if needed. The SPLENDAQ package provides a tutorial on GitHub~\cite{splendaq} on converting raw data into the required format. With the data in this format, the data triggering algorithms can be run and the extracted events will have the same format.

\begin{figure}
    \centering
    \includegraphics[width=0.4\linewidth]{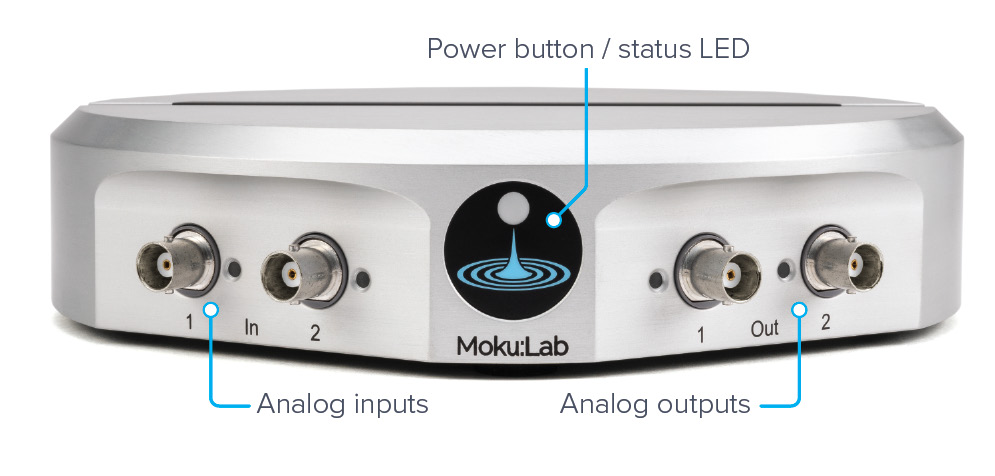}
    \includegraphics[width=0.4\linewidth]{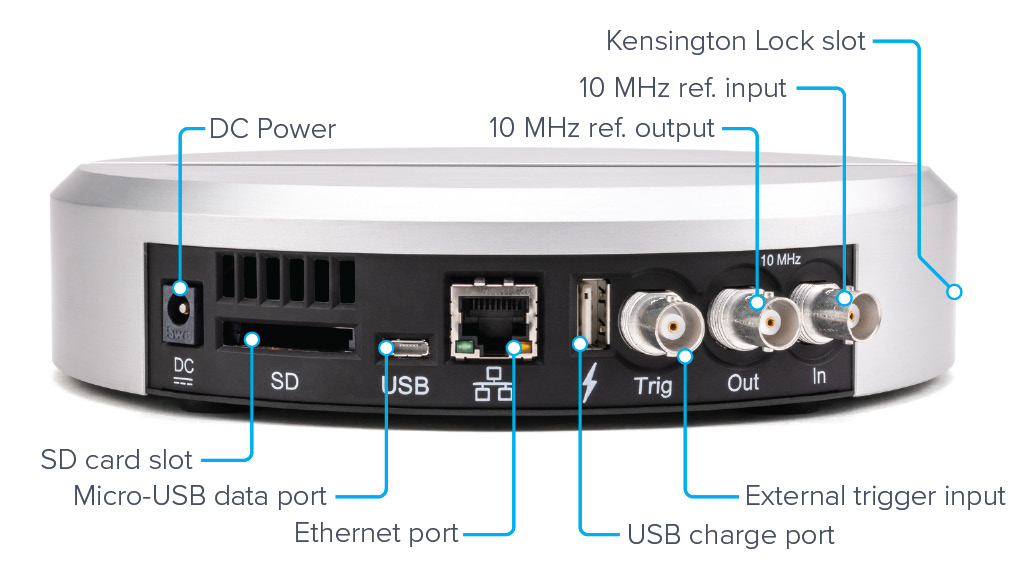}
    \caption{From the Moku:Lab technical specifications provided by Liquid Instruments, we show the front and back of the Moku:Lab and label the various ports. We refer the reader to the Liquid Instruments website~\cite{liquidinstruments} for further information on the specifications and the other Moku models.}
    \label{fig:mokulab}
\end{figure}

\textit{Example usage.---}While the described triggering algorithms do not require a specific system for logging the continuous data, we have been using the Moku platform, developed by Liquid Instruments~\cite{liquidinstruments}, in order to acquire and save a continuous data stream for offline processing with SPLENDAQ. There are three models that have been developed by and are available for purchase from Liquid Instruments: the Moku:Pro, Moku:Lab, and Moku:Go. The SPLENDAQ package provides helper functions for interfacing with each of these models for ease of use, as Liquid Instruments provides various application programming interfaces (APIs) for interacting with the Moku. In this work, we will depict some example usage of the data triggering algorithms using the Moku:Lab platform (as pictured in Fig.~\ref{fig:mokulab}), knowing this usage would have identical steps for each model. We note that there are many more features to the Moku than continuous data logging, but we focus on this one feature to provide a simple depiction of using SPLENDAQ for offline triggering.

\begin{figure}
    \centering
    \includegraphics[width=1.0\linewidth]{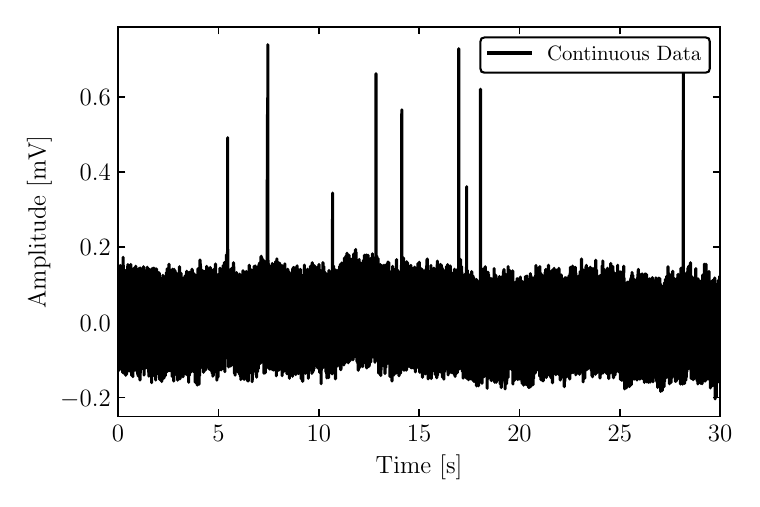}
    \caption{A section of the continuous data taken with the Moku device, where the pulses we added to the data are seen as spikes at this level of zoom.}
    \label{fig:cont_data}
\end{figure}

With the Moku, we can log a continuous stream of data, where here we simply read the intrinsic noise of the system. Here, we take $30\,\mathrm{s}$ of noise data at a digitization rate of $250\,\mathrm{kHz}$, specifying a peak-to-peak voltage range of $1\,\mathrm{V}$ (note that the ADC resolution of the Moku:Lab is 12 bits). There is technically no maximum length of the continuous data, but keeping each continuous data file on the order of a minute is good practice to avoid RAM issues when loading the logged data. The Moku itself saves binary files (with file extension \texttt{.li}) with a specific encoding defined by Liquid Instruments, this file being of size 30\,MB. SPLENDAQ provides a helper function called \texttt{splendaq.io.convert\_li\_to\_h5}, which converts \texttt{.li} files to the HDF5-based data format required by SPLENDAQ for triggering. After converting the file, we can add ``fake" pulses to the data by defining a known template for this demonstration of the triggering functionality. Specifically, our data has 10 double-exponential pulses each with a characteristic rise time of $20\,\mu\mathrm{s}$ and a characteristic fall time of $58 \,\mu\mathrm{s}$, randomly spaced with in the data stream, as pictured in Fig~\ref{fig:cont_data}. As we know the pulse shape a priori, we will use a signal template with the same characteristic time constants. The heights of these pulses were generated from a normal distribution with mean $0.5\,\mathrm{mV}$ and standard deviation $0.125 \, \mathrm{mV}$, which corresponds to amplitudes of about 35$\sigma_A$ (we evaluate $\sigma_A$ in the next paragraph).

\begin{figure}
    \centering
    \includegraphics[width=1.0\linewidth]{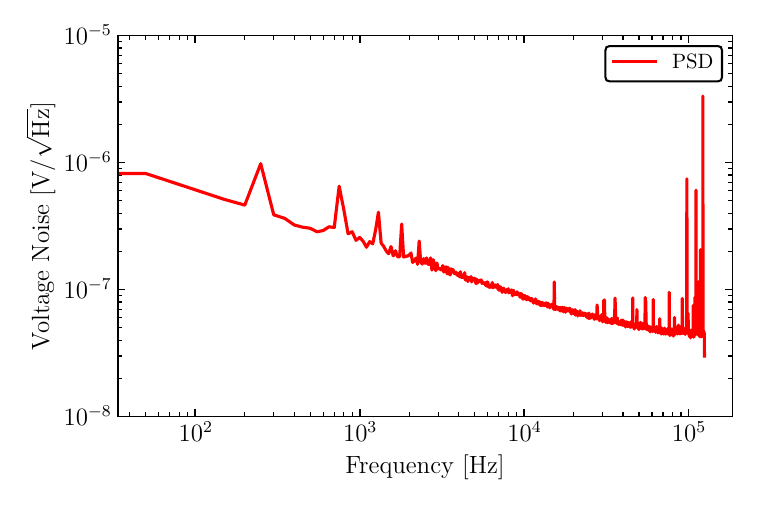}
    \caption{The measured intrinsic voltage noise of the continuous data after removing events from the population of randoms that have been contaminated by pulses (i.e. after applying ``cuts").}
    \label{fig:psd}
\end{figure}

The SPLENDAQ class that facilitates using the data triggering algorithms is \texttt{splendaq.daq.EventBuilder}, which requires the path to a folder that contains the continuous data, a path to a folder to save the triggered data, and the length (in time bins) of the waveforms to save. The signal template should match this specified length. Here, we stipulate that the waveforms should be $l=5000$ bins long, equivalent to $20\,\mathrm{ms}$ at the $250\,\mathrm{kHz}$ digitization rate of this data. With the class initialized, we use the class method \texttt{EventBuilder.acquire\_randoms} class method to run the random triggering algorithm. With a desired waveform length of $20\,\mathrm{ms}$, this means that the maximum amount of nonoverlapping random windows that can be saved from the $30\,\mathrm{s}$ data is 1500. To ensure we do not approach this limit (which would give essentially the continuous data split into 1500 chunks without any spacing between the chunks), we instead save $n=500$ waveforms. After running the random triggering algorithm, the corresponding files are saved in the specified folder for storing triggered data, a step which takes about 0.8\,s  and creates a 96\,MB file for this dataset on a personal computer. As we need to calculate the noise PSD to be used for the optimal filter, we can open these files, clean the data of any pulses using some type of data selection algorithm, and calculate the PSD with the remaining events. To open the files, SPLENDAQ supplies a class called \texttt{splendaq.io.Reader}. Following this prescription, we show the voltage noise PSD in Fig.~\ref{fig:psd}, where we can see the measured intrinsic noise after removing events contaminated by pulses. Using Eq.~(\ref{eq:of_res}), we find that the baseline resolution of the system is $\sigma_A = 14\,\mu\mathrm{V}$ via our measured PSD and known signal template.

\begin{table*}
    \caption{\label{tab:values}The expected and measured amplitudes for each of the 11 detected pulses. For the known pulses, there is excellent agreement between the expected and measured pulses heights (i.e. amplitudes), given the expected signal resolution of $\sigma_A=14\,\mu\mathrm{V}$. Note that $A_4$ has a dash for the expected height, as this was a noise triggered event that had a height that was just above the threshold of $5\sigma_A=69 \, \mu\mathrm{V}$.}
    \begin{ruledtabular}
        \begin{tabular}{crrrrrrrrrrr}
          Amplitude [$\mu\mathrm{V}$] & $A_1$ & $A_2$ & $A_3$ & $A_4$ & $A_5$ & $A_6$ & $A_7$ & $A_8$ & $A_9$ & $A_{10}$ & $A_{11}$\\ \hline
          Expected     & 413 & 474 & 666 & ---   & 308 & 614 & 548 & 696  & 314 & 600 & 715\\
          Measured  & 403 & 480 & 677 & 70 & 294 & 594 &  542 & 702 & 314  & 581 &  729 \\ 
        \end{tabular}
    \end{ruledtabular}
\end{table*}

\begin{figure}
    \centering
    \includegraphics[width=1.0\linewidth]{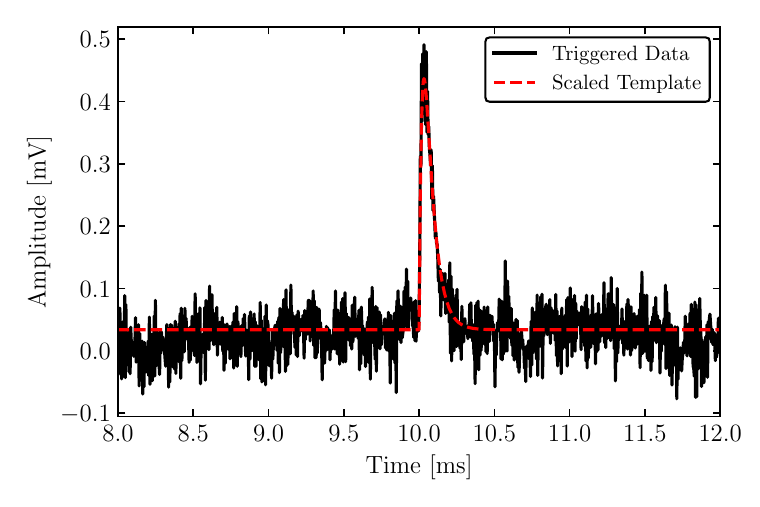}
    \caption{An event that was acquired from the continuous data stream using the optimal filter--based threshold triggering algorithm, which corresponds to the $A_1$ in Table~\ref{tab:values}. We also show the signal template scaled to the extracted amplitude and adjusted to match the DC baseline of the data. Note the excellent agreement between the two, showing that we have reconstructed the amplitude of the pulse well. For this fit, $\chi^2=4938\pm99$, which is consistent with the expected value of 5000.}
    \label{fig:trigger}
\end{figure}

With a signal template and a noise PSD, we can now use the optimal filter--based algorithm to acquire threshold-triggered data. The SPLENDAQ class method \texttt{EventBuilder.acquire\_pulses} facilitates using the triggering algorithm, which requires the signal template, the noise PSD, the threshold in number of $\sigma_A$, and the index of which channel to run the algorithm on (for the data in this work, there is a single channel, so this is set to 0). For this data, we set the threshold to $5\sigma_A$, and we set the method's keyword argument \texttt{mergewindow=2500} (in number of time bins) to merge events that are within half of a waveform length. We then run the threshold triggering algorithm (this takes about 0.7\,s on a personal computer and creates a 2.4\,MB file), which returns 11 events, indicating that there was a noise trigger. In Table~\ref{tab:values}, we show the expected and measured pulse heights of each of these events. For each pulse added, the reconstructed energy is within 1 or 2 $\sigma_A$ of the true value, showing that the reconstruction algorithm is operating as expected. In Fig.~\ref{fig:trigger}, we show one of these events, comparing the triggered event to the template scaled to the reconstructed amplitude of the event. From this, we see that the threshold triggering algorithm reconstructed the pulse accurately, extracting the expected $0.4\,\mathrm{mV}$ amplitude of the pulse.

\textit{Conclusion.---}With SPLENDAQ, we provide easy-to-use functionality for running data triggering algorithms on a continuous data stream, both a random triggering algorithm and threshold triggering algorithm. In this work, we have described the underlying algorithms, specified the data format requirements, and shown a simple example of the workflow. As this package has been designed to be simple to set up and to be agnostic to  the source of data, the use cases can span across many scientific disciplines. As of now, it has been used for characterization of various types of detectors, materials measurements, and assays of radioactive materials. We will continue to improve the functionality of and add new algorithms to SPLENDAQ, including plans to allow for multi-channel triggering, merging of coincidence triggers between channels, and support for more hardware platforms beyond the Moku. As the package is on GitHub, we recommend any interested users to try out the tutorials and welcome collaborators from all technical backgrounds to contribute to the codebase.

\textit{Acknowledgments.---}This work was supported by the U.S. Department of Energy through the Los Alamos National Laboratory. Los Alamos National Laboratory is operated by Triad National Security, LLC, for the National Nuclear Security Administration of U.S. Department of Energy (Contract No. 89233218CNA000001). Research presented in this article was supported by the Laboratory Directed Research and Development program of Los Alamos National Laboratory under project numbers 20220135DR and 20230782PRD1.


\begin{thebibliography}{14}%
\makeatletter
\providecommand \@ifxundefined [1]{%
 \@ifx{#1\undefined}
}%
\providecommand \@ifnum [1]{%
 \ifnum #1\expandafter \@firstoftwo
 \else \expandafter \@secondoftwo
 \fi
}%
\providecommand \@ifx [1]{%
 \ifx #1\expandafter \@firstoftwo
 \else \expandafter \@secondoftwo
 \fi
}%
\providecommand \natexlab [1]{#1}%
\providecommand \enquote  [1]{``#1''}%
\providecommand \bibnamefont  [1]{#1}%
\providecommand \bibfnamefont [1]{#1}%
\providecommand \citenamefont [1]{#1}%
\providecommand \href@noop [0]{\@secondoftwo}%
\providecommand \href [0]{\begingroup \@sanitize@url \@href}%
\providecommand \@href[1]{\@@startlink{#1}\@@href}%
\providecommand \@@href[1]{\endgroup#1\@@endlink}%
\providecommand \@sanitize@url [0]{\catcode `\\12\catcode `\$12\catcode `\&12\catcode `\#12\catcode `\^12\catcode `\_12\catcode `\%12\relax}%
\providecommand \@@startlink[1]{}%
\providecommand \@@endlink[0]{}%
\providecommand \url  [0]{\begingroup\@sanitize@url \@url }%
\providecommand \@url [1]{\endgroup\@href {#1}{\urlprefix }}%
\providecommand \urlprefix  [0]{URL }%
\providecommand \Eprint [0]{\href }%
\providecommand \doibase [0]{https://doi.org/}%
\providecommand \selectlanguage [0]{\@gobble}%
\providecommand \bibinfo  [0]{\@secondoftwo}%
\providecommand \bibfield  [0]{\@secondoftwo}%
\providecommand \translation [1]{[#1]}%
\providecommand \BibitemOpen [0]{}%
\providecommand \bibitemStop [0]{}%
\providecommand \bibitemNoStop [0]{.\EOS\space}%
\providecommand \EOS [0]{\spacefactor3000\relax}%
\providecommand \BibitemShut  [1]{\csname bibitem#1\endcsname}%
\let\auto@bib@innerbib\@empty
\bibitem [{\citenamefont {Howe}\ \emph {et~al.}(2004)\citenamefont {Howe}, \citenamefont {Cox}, \citenamefont {Harvey}, \citenamefont {McGirt}, \citenamefont {Rielage}, \citenamefont {Wilkerson},\ and\ \citenamefont {Wouters}}]{orca2004}%
  \BibitemOpen
  \bibfield  {author} {\bibinfo {author} {\bibfnamefont {M.~A.}\ \bibnamefont {Howe}}, \bibinfo {author} {\bibfnamefont {G.~A.}\ \bibnamefont {Cox}}, \bibinfo {author} {\bibfnamefont {P.~J.}\ \bibnamefont {Harvey}}, \bibinfo {author} {\bibfnamefont {F.}~\bibnamefont {McGirt}}, \bibinfo {author} {\bibfnamefont {K.}~\bibnamefont {Rielage}}, \bibinfo {author} {\bibfnamefont {J.~F.}\ \bibnamefont {Wilkerson}},\ and\ \bibinfo {author} {\bibfnamefont {J.~M.}\ \bibnamefont {Wouters}},\ }\bibfield  {title} {\bibinfo {title} {Sudbury neutrino observatory neutral current detector acquisition software overview},\ }\href {https://doi.org/10.1109/TNS.2004.829527} {\bibfield  {journal} {\bibinfo  {journal} {IEEE Trans. Nucl. Sci.}\ }\textbf {\bibinfo {volume} {51}},\ \bibinfo {pages} {878} (\bibinfo {year} {2004})}\BibitemShut {NoStop}%
\bibitem [{\citenamefont {Khachatryan}\ \emph {et~al.}(2017)\citenamefont {Khachatryan} \emph {et~al.}}]{CMS:2016ngn}%
  \BibitemOpen
  \bibfield  {author} {\bibinfo {author} {\bibfnamefont {V.}~\bibnamefont {Khachatryan}} \emph {et~al.} (\bibinfo {collaboration} {CMS Collaboration}),\ }\bibfield  {title} {\bibinfo {title} {{The CMS trigger system}},\ }\href {https://doi.org/10.1088/1748-0221/12/01/P01020} {\bibfield  {journal} {\bibinfo  {journal} {J. Instrum.}\ }\textbf {\bibinfo {volume} {12}}\bibfield  {number} {\bibinfo  {number} { (01)},\ \bibinfo {pages} {P01020}},\ }\Eprint {https://arxiv.org/abs/1609.02366} {arXiv:1609.02366 [physics.ins-det]} \BibitemShut {NoStop}%
\bibitem [{\citenamefont {Di~Domizio}\ \emph {et~al.}(2018)\citenamefont {Di~Domizio}, \citenamefont {Branca}, \citenamefont {Caminata}, \citenamefont {Canonica}, \citenamefont {Copello}, \citenamefont {Giachero}, \citenamefont {Guardincerri}, \citenamefont {Marini}, \citenamefont {Pallavicini},\ and\ \citenamefont {Vignati}}]{DiDomizio:2018ldc}%
  \BibitemOpen
  \bibfield  {author} {\bibinfo {author} {\bibfnamefont {S.}~\bibnamefont {Di~Domizio}}, \bibinfo {author} {\bibfnamefont {A.}~\bibnamefont {Branca}}, \bibinfo {author} {\bibfnamefont {A.}~\bibnamefont {Caminata}}, \bibinfo {author} {\bibfnamefont {L.}~\bibnamefont {Canonica}}, \bibinfo {author} {\bibfnamefont {S.}~\bibnamefont {Copello}}, \bibinfo {author} {\bibfnamefont {A.}~\bibnamefont {Giachero}}, \bibinfo {author} {\bibfnamefont {E.}~\bibnamefont {Guardincerri}}, \bibinfo {author} {\bibfnamefont {L.}~\bibnamefont {Marini}}, \bibinfo {author} {\bibfnamefont {M.}~\bibnamefont {Pallavicini}},\ and\ \bibinfo {author} {\bibfnamefont {M.}~\bibnamefont {Vignati}},\ }\bibfield  {title} {\bibinfo {title} {{A data acquisition and control system for large mass bolometer arrays}},\ }\href {https://doi.org/10.1088/1748-0221/13/12/P12003} {\bibfield  {journal} {\bibinfo  {journal} {J. Instrum.}\ }\textbf {\bibinfo {volume} {13}}\bibfield  {number} {\bibinfo  {number} { (12)},\ \bibinfo {pages} {P12003}},\ }\Eprint
  {https://arxiv.org/abs/1807.11446} {arXiv:1807.11446 [physics.ins-det]} \BibitemShut {NoStop}%
\bibitem [{\citenamefont {Wilson}\ \emph {et~al.}(2022)\citenamefont {Wilson} \emph {et~al.}}]{Wilson:2022quj}%
  \BibitemOpen
  \bibfield  {author} {\bibinfo {author} {\bibfnamefont {J.~S.}\ \bibnamefont {Wilson}} \emph {et~al.},\ }\bibfield  {title} {\bibinfo {title} {{The level-1 trigger for the SuperCDMS experiment at SNOLAB}},\ }\href {https://doi.org/10.1088/1748-0221/17/07/P07010} {\bibfield  {journal} {\bibinfo  {journal} {J. Instrum.}\ }\textbf {\bibinfo {volume} {17}}\bibfield  {number} {\bibinfo  {number} { (07)},\ \bibinfo {pages} {P07010}},\ }\Eprint {https://arxiv.org/abs/2204.13002} {arXiv:2204.13002 [physics.ins-det]} \BibitemShut {NoStop}%
\bibitem [{\citenamefont {Watkins}(2023)}]{splendaq}%
  \BibitemOpen
  \bibfield  {author} {\bibinfo {author} {\bibfnamefont {S.~L.}\ \bibnamefont {Watkins}} (\bibinfo {collaboration} {SPLENDOR Collaboration}),\ }\href@noop {} {\bibinfo {title} {{SPLENDAQ}}} (\bibinfo {year} {2023}),\ \bibinfo {note} {\href{https://www.github.com/splendor-collab/splendaq}{https://www.github.com/splendor-collab/splendaq}}\BibitemShut {NoStop}%
\bibitem [{\citenamefont {Jordanov}\ \emph {et~al.}(1994)\citenamefont {Jordanov}, \citenamefont {Knoll}, \citenamefont {Huber},\ and\ \citenamefont {Pantazis}}]{JORDANOV1994261}%
  \BibitemOpen
  \bibfield  {author} {\bibinfo {author} {\bibfnamefont {V.~T.}\ \bibnamefont {Jordanov}}, \bibinfo {author} {\bibfnamefont {G.~F.}\ \bibnamefont {Knoll}}, \bibinfo {author} {\bibfnamefont {A.~C.}\ \bibnamefont {Huber}},\ and\ \bibinfo {author} {\bibfnamefont {J.~A.}\ \bibnamefont {Pantazis}},\ }\bibfield  {title} {\bibinfo {title} {Digital techniques for real-time pulse shaping in radiation measurements},\ }\href {https://doi.org/10.1016/0168-9002(94)91652-7} {\bibfield  {journal} {\bibinfo  {journal} {Nucl. Instrum. Meth. Phys. Res. A}\ }\textbf {\bibinfo {volume} {353}},\ \bibinfo {pages} {261} (\bibinfo {year} {1994})}\BibitemShut {NoStop}%
\bibitem [{\citenamefont {{Radeka}}\ and\ \citenamefont {{Karlovac}}(1967)}]{radeka1967}%
  \BibitemOpen
  \bibfield  {author} {\bibinfo {author} {\bibfnamefont {V.}~\bibnamefont {{Radeka}}}\ and\ \bibinfo {author} {\bibfnamefont {N.}~\bibnamefont {{Karlovac}}},\ }\bibfield  {title} {\bibinfo {title} {{Least-square-error amplitude measurement of pulse signals in presence of noise}},\ }\href {https://doi.org/10.1016/0029-554X(67)90561-7} {\bibfield  {journal} {\bibinfo  {journal} {Nucl. Instrum. Meth.}\ }\textbf {\bibinfo {volume} {52}},\ \bibinfo {pages} {86} (\bibinfo {year} {1967})}\BibitemShut {NoStop}%
\bibitem [{\citenamefont {Gatti}\ and\ \citenamefont {Manfredi}(1986)}]{Gatti:1986cw}%
  \BibitemOpen
  \bibfield  {author} {\bibinfo {author} {\bibfnamefont {E.}~\bibnamefont {Gatti}}\ and\ \bibinfo {author} {\bibfnamefont {P.~F.}\ \bibnamefont {Manfredi}},\ }\bibfield  {title} {\bibinfo {title} {{Processing the Signals From Solid State Detectors in Elementary Particle Physics}},\ }\href {https://doi.org/10.1007/BF02822156} {\bibfield  {journal} {\bibinfo  {journal} {Riv. Nuovo Cim.}\ }\textbf {\bibinfo {volume} {9N1}},\ \bibinfo {pages} {1} (\bibinfo {year} {1986})}\BibitemShut {NoStop}%
\bibitem [{\citenamefont {Golwala}(2000)}]{golwala}%
  \BibitemOpen
  \bibfield  {author} {\bibinfo {author} {\bibfnamefont {S.~R.}\ \bibnamefont {Golwala}},\ }\emph {\bibinfo {title} {Exclusion limits on the WIMP nucleon elastic scattering cross-section from the Cryogenic Dark Matter Search}},\ \href {https://doi.org/10.2172/1421437} {Ph.D. thesis},\ \bibinfo  {school} {University of California, Berkeley} (\bibinfo {year} {2000})\BibitemShut {NoStop}%
\bibitem [{\citenamefont {Watkins}(2022)}]{Watkins:2022wlz}%
  \BibitemOpen
  \bibfield  {author} {\bibinfo {author} {\bibfnamefont {S.~L.}\ \bibnamefont {Watkins}},\ }\emph {\bibinfo {title} {{Athermal Phonon Sensors in Searches for Light Dark Matter}}},\ \href@noop {} {Ph.D. thesis},\ \bibinfo  {school} {University of California, Berkeley} (\bibinfo {year} {2022}),\ \Eprint {https://arxiv.org/abs/2301.08699} {arXiv:2301.08699 [hep-ex]} \BibitemShut {NoStop}%
\bibitem [{\citenamefont {Harris}\ \emph {et~al.}(2020)\citenamefont {Harris} \emph {et~al.}}]{harris2020array}%
  \BibitemOpen
  \bibfield  {author} {\bibinfo {author} {\bibfnamefont {C.~R.}\ \bibnamefont {Harris}} \emph {et~al.},\ }\bibfield  {title} {\bibinfo {title} {Array programming with {NumPy}},\ }\href {https://doi.org/10.1038/s41586-020-2649-2} {\bibfield  {journal} {\bibinfo  {journal} {Nature}\ }\textbf {\bibinfo {volume} {585}},\ \bibinfo {pages} {357} (\bibinfo {year} {2020})},\ \Eprint {https://arxiv.org/abs/2006.10256} {arXiv:2006.10256 [cs.MS]} \BibitemShut {NoStop}%
\bibitem [{\citenamefont {Virtanen}\ \emph {et~al.}(2020)\citenamefont {Virtanen} \emph {et~al.}}]{2020SciPy-NMeth}%
  \BibitemOpen
  \bibfield  {author} {\bibinfo {author} {\bibfnamefont {P.}~\bibnamefont {Virtanen}} \emph {et~al.},\ }\bibfield  {title} {\bibinfo {title} {{{SciPy} 1.0: Fundamental Algorithms for Scientific Computing in Python}},\ }\href {https://doi.org/10.1038/s41592-019-0686-2} {\bibfield  {journal} {\bibinfo  {journal} {Nat. Methods}\ }\textbf {\bibinfo {volume} {17}},\ \bibinfo {pages} {261} (\bibinfo {year} {2020})},\ \Eprint {https://arxiv.org/abs/1907.10121} {arXiv:1907.10121 [cs.MS]} \BibitemShut {NoStop}%
\bibitem [{\citenamefont {{The HDF Group}}(2023)}]{hdf5}%
  \BibitemOpen
  \bibfield  {author} {\bibinfo {author} {\bibnamefont {{The HDF Group}}},\ }\href@noop {} {\bibinfo {title} {{Hierarchical Data Format, version 5}}} (\bibinfo {year} {1997-2023}),\ \bibinfo {note} {\href{https://www.hdfgroup.org/HDF5/}{https://www.hdfgroup.org/HDF5/}}\BibitemShut {NoStop}%
\bibitem [{\citenamefont {{Liquid Instruments}}(2023)}]{liquidinstruments}%
  \BibitemOpen
  \bibfield  {author} {\bibinfo {author} {\bibnamefont {{Liquid Instruments}}},\ }\href@noop {} {\bibinfo {title} {{Moku Hardware Platforms}}} (\bibinfo {year} {2023}),\ \bibinfo {note} {\href{https://www.liquidinstruments.com/}{https://www.liquidinstruments.com/}}\BibitemShut {NoStop}%
\end{thebibliography}
\end{document}